\documentclass[preprint]{ptephy_v1}%%%%where ptephy_v1 is the template name

\preprintnumber{XXXX-XXXX} %%% %%% Insert preprint number here

\usepackage{graphicx}
\usepackage[T1]{fontenc}
\usepackage{multirow}
\usepackage{tabularx}
\usepackage{xcolor}
\usepackage{threeparttable}
\usepackage{bm}
\usepackage{hyperref}
\hypersetup{
    breaklinks,
    unicode,
    linktoc=all,
    bookmarksnumbered=true,
    bookmarksopen=true,
    colorlinks,
    linkcolor=blue,
    citecolor=blue,
    urlcolor=blue,
    plainpages=false,
    pdfstartview=FitH,
    pdfborder={0 0 0},
    linktocpage
}

\newcommand{\ts}{\textsuperscript}
\newcommand{\tb}{\textsubscript}
\begin{document}

\title{Shell Migration at \textbf{\textit{N}} = 32, 34 around Ca Region}

%\newcommand{\textsc}

%%%% To generate auto affiliation numbers please use \author{}\affil{} command

\author{Hongna Liu}
\affil{Key Laboratory of Beam Technology of Ministry of Education, School of Physics and Astronomy,
Beijing Normal University, Beijing 100875, China \email{hongna.liu@bnu.edu.cn}} 

\author{Sidong Chen}
\affil{School of Physics, Engineering and Technology, University of York, Heslington, York YO10 5DD, UK, \email{sidong.chen@york.ac.uk}}

\author{Frank Browne}
\affil{Department of Physics and Astronomy, University of Manchester, Manchester, M13 9PL, United Kingdom, \email{frank.browne@manchester.ac.uk}}

\begin{abstract}
Shell closures at $\emph{N}$ = 32 and 34, not present in stable nuclei, have been suggested in neutron-rich pf-shell nuclei. In this article, we discuss the experimental observables and state-of-the-art theoretical calculations that characterize and explain the shell evolution leading to new magic numbers. Particular focus shall be afforded to the experimental progress of the shell migration study at and beyond $\emph{N}$ = 32, 34 in Ar, K, Ca, and Sc isotopes at the RIBF using direct reactions with liquid hydrogen targets over the past ten years. The results prove the double magicity of \ts{52,54}Ca, and support the persistence of the $\emph{N}$ = 34 subshell closure below $\emph{Z}$ = 20 with a sharp weakening beyond $\emph{Z}$ = 20.  Future measurements of intruder bands of $\emph{N}$ = 32, 34 nuclei and shell evolution towards $\emph{N}$ = 40 are discussed within the context of an upgraded RIBF facility and the development of novel detection systems. 
\end{abstract}

\subjectindex{xxxx, xxx}

%\pacs{xxxxxxxx}
\maketitle

\section{Introduction}
Achieving a universal understanding of the underlying forces from which nuclear structure emerges is a primary goal of the nuclear physics field. The canonical nuclear magic numbers, which represent particularly stable configurations, were observed at 2, 8, 20, 28, 50, 82, and 126~\cite{mayer_pr_1948}, and reproduced through the description of protons and neutrons bound in a self-induced attractive mean field with a strong spin-orbit ($LS$) interaction~\cite{haxel_pr_1949,mayer_pr_1949}. 

Through the development of radioactive beam technology, extensive studies have been made on the evolution of shell structures through the systematic behaviour of properties of isotopic and isotonic chains, where Fermi surfaces of protons and neutrons, respectively, are constant. Magic numbers established for nuclei close to beta-stability are found not to be universal across the nuclear landscape. For example, the conventional $\emph{N}$ = 8, 20, 28 magic numbers disappear and new magic numbers, such as $\emph{N}$ = 16, emerge in neutron-rich nuclei.  Such phenomena offer new insights into the underlying nuclear forces that drive these structural changes. A comprehensive review of shell evolution can be found in Ref.~\cite{otsuka_rmp_2020}. This article focuses on the shell evolution in the neutron-rich Ca region, where $N=32,34$ have been suggested as new magic numbers from calculations and previous measurements~\cite{Huck1985,Gade2006,steppenbeck_nature_2013,wienholtz_nature_2013}. We will review key experimental results obtained using the quasi-free scattering from liquid hydrogen targets at the RIBF and discuss future perspectives in the neutron-rich Ca region.
%FB commended this bit out, feels out of place.... 
%The origin of the shell evolution along isotones was attributed to the change of spin-orbit splitting due to the tensor force, while three-body forces that are mainly repulsive to avoid overbinding of the nuclei was suggested to play an important role for the evolution along isotopes. \textcolor{red}{Need to include some references. FB.}\\

The Ca isotopes are defined by their robust $\emph{Z}$ = 20 proton shell closure, providing a good environment to study the neutron shell migration along the isotopic chain. Such sensitivity has been instrumental in guiding the development of so-called effective interactions, which take inputs from experiments and derive the parameters of the associated interactions. In addition, the Ca isotopic chain is accessible by calculations from a variety of theoretical approaches such as mean-field, large-scale shell model and {\em ab initio} approaches. Experimental information in the Ca region is thus highly desirable to test the cutting-edge nuclear structure theories and is critical to pin down the driving forces behind the shell evolution.

%In the neutron-rich $pf$-shell nuclei, possible new magic numbers $\emph{N}$ = 32~\cite{Huck1985,wienholtz_nature_2013,Gade2006}, $\emph{N}$ = 34~\cite{steppenbeck_nature_2013,michimasa_prl_2018,chen_prl_2019} have been suggested by calculations. % FB redoing this bit to only cite calculations... 
In neutron-rich $pf$-shell nuclei, possible new magic numbers $\emph{N}$ = 32 and $\emph{N}$ = 34 have been suggested qualitatively through calculations, for example in Refs.~\cite{otsuka_prl_2001,rodriguez_prl_2007,honma_rapr_2008,coraggio_prc_2009}. In the shell model framework, it suggests that the attractive tensor interactions between protons in ${\pi}0f_{7/2}$ orbitals and neutrons in ${\nu}0f_{5/2}$ orbitals decreases from $^{64}$Ni to $^{54}$Ca, leading to a continuous increase of the energy of the ${\nu}0f_{5/2}$ orbitals with respect to the ${\nu}1p_{3/2}$--${\nu}1p_{1/2}$. Here forms the $\emph{N}$ = 34 subshell closure lying above the $N=32$ subshell, which itself is caused by the respulsion of the $1p_{3/2}$ and $1p_{1/2}$ orbitals~\cite{holt_jpg_2012}.
Theoretical calculations based on $\emph{ab initio}$ coupled cluster theory with realistic nuclear interactions have shown that three-body (3N) forces are crucial to correctly describe the structure of neutron-rich Ca isotopes~\cite{holt_jpg_2012}. Experimental data on shell migration at $\emph{N}$ = 32 and 34 is necessary to verify those predicted pictures. In addition, $^{60}$Ca is a candidate as the last double-$\emph{LS}$ magic nucleus, following $^{4}$He, $^{16}$O and $^{40}$Ca~\cite{Brown2022PHY}. It has been discovered that $^{60}$Ca lies inside the neutron drip line~\cite{Tarasov2018PRL}, but its mass and excited states are still inaccessible by experiment.

Beyond $N=32$, the spectroscopic study of neutron-rich Ca isotopes has been limited by shortcomings in luminosities afforded by accelerator facilities, indeed, following the first spectroscopy of \ts{52}Ca in 1985~\cite{Huck1985}, it took over 30 years before \ts{54}Ca could be studied~\cite{steppenbeck_nature_2013}. For a given facility, the luminosity, which is a combined effect of the beam intensity, cross sections of the selected reaction process, and the allowed target thickness, determines its accessible range of exotic nuclei. 
Although beam intensity generally improves over the lifespan of a facility, it remains limited. Therefore, selecting reaction mechanisms that maximize cross sections and usable target thickness is essential for reaching evermore exotic nuclei. In-beam $\gamma$-ray spectroscopy following quasi-free scattering with a hydrogen target in inverse kinematics at intermediate energies has been proven to be an effective method to study the microscopic structure of atomic nuclei. In the case of $(p,pN)$, it features a reasonably large cross section from a few to tens of millibarns~\cite{paul_prl_2019}. In $(p,pN)$ reaction at the RIBF energy in inverse kinematics, the scattered proton and the struck nucleon from $(p,pN)$ are emitted following a large momentum transfer with a relative emission angle centered about 40 degrees. Through tracking the target recoil proton and, if applicable, the knocked out proton the reaction vertex can be reconstructed, allowing the use of very thick targets to improve the luminosity without degradation of information important to the analysis of the reaction and its subsequent decays. The MINOS device, comprising a liquid hydrogen target with a length up to 150\,mm and a Time Projection Chamber (TPC) proton tracker, pioneered the exploitation of this principle in the application to nuclear structure and reaction studies~\cite{obertelli_epja_2014}. Conceived for in-beam $\gamma$ spectroscopy of bound states and invariant mass spectroscopy of unbound states, the MINOS device was coupled with the NaI(Tl) $\gamma$-ray spectrometer DALI2 at the RIBF~\cite{takeuchi_nima_2014} for the SEASTAR campaigns~\cite{seastar}, of which the third campaign conducted in 2017 aimed at spectroscopy of the very neutron-rich Ca region. Compared to a typically-used passive heavy-ion target, such as \ts{9}Be of 10~mm thickness, MINOS offers a gain in luminosity by a factor of $\sim5$--10 with the same energy resolution and allows in-beam $\gamma$-ray spectroscopy of nuclei with a production rate down to 1 particle per second (pps).

In this article, the experimental observables that characterize the shell evolution are given in Section 2, and then, Section 3 provides an overview of theoretical approaches used to study shell migration in neutron-rich $pf$-shell region. In Section 4, the status of shell migration at $\emph{N}$ = 32, 34 and towards $\emph{N}$ = 40 in the Ca region is presented with a focus on the experimental progress achieved using the MINOS and DALI2 setups at the RIBF facility in the past ten years. 
%In Section 5, we discuss the perspective about shell evolution study in Ca region with the newly constructed devices and the upgrade RIBF

\section{Experimental observables of shell evolution}
%To characterize shell evolution, we need to find sensitive probes, however, t

\begin{figure}[bth]
\centering
\includegraphics[width=1.02\textwidth]{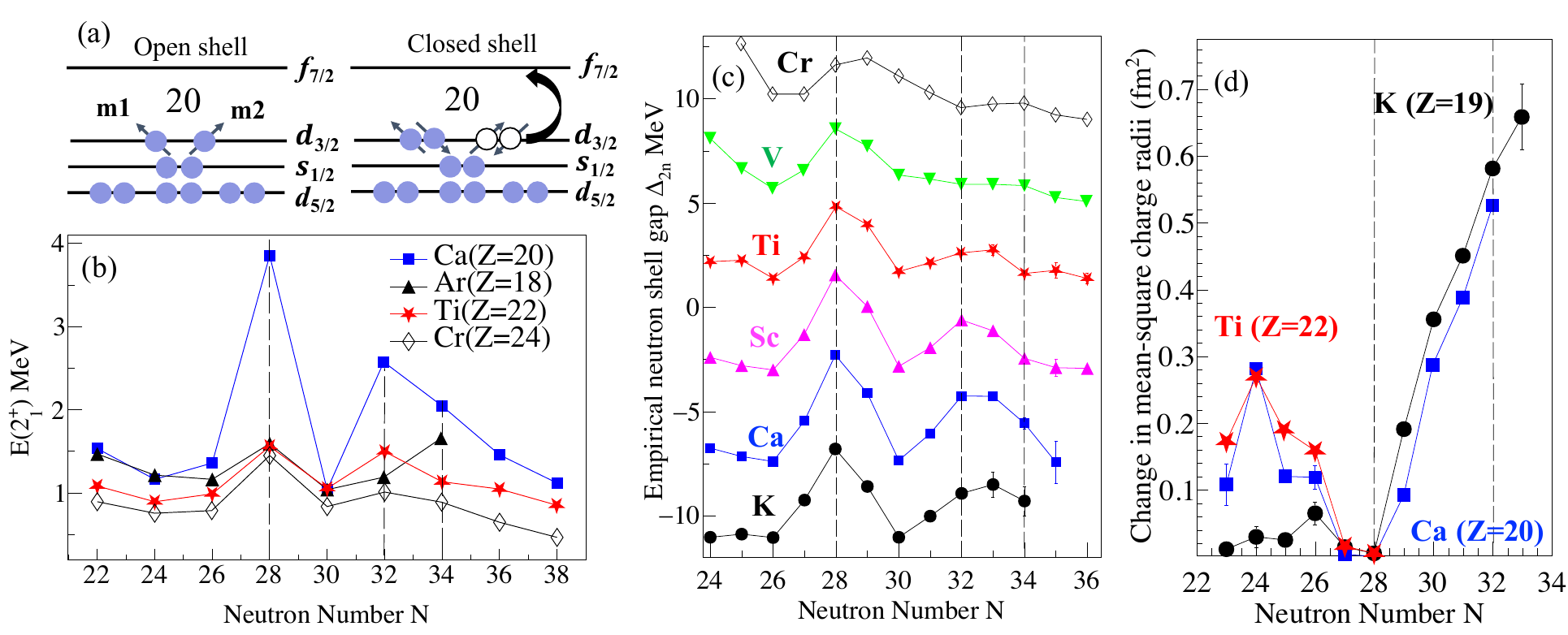}
    \caption{(a) Formation of $2_{1}^{+}$ for open and closed shell nuclei. (b) Systematics of the measured $E(2^+_1)$ for $pf$-shell neutron-rich nuclei in Ca region. Data are taken from Ref.~\cite{nndc}. (c) Deduced empirical neutron-shell gaps ($\Delta_{2n}$) for K, Ca, Sc, Ti, V, and Cr isotopes \cite{wienholtz_nature_2013,Rosenbusch2015PRL,Leistenschneider2018PRL, Leistenschneider2021, Iimura2023}. The Ti isotopes remain in their original positions, while the other isotopic chains have been shifted by multiples of 4 MeV for clarity. (d) Changes of the mean-square charge radii from K to Fe, relative to isotope with neutron number $\emph{N}$ = 28. Data are taken from Ref.~\cite{Koszorus2021}.}
    \label{fig:e2_charge}
\end{figure}

The shell gap, defined as the difference between the effective single-particle energies of the orbitals, is not an observable but sensitive to several measurable properties of atomic nuclei. By integrating information from different facets, we will be able to draw a portrait of the nuclei that closely reflects its reality. 

In a simplified shell model picture, for closed-shell even-even nuclei shown in Fig.~\ref{fig:e2_charge} (a), the formation of $2_{1}^{+}$ state requires cross-shell excitations since the valence orbitals are fully occupied, while for open-shell nuclei, it only involves the rearrangement of nucleons within the valence space. Therefore, a sudden increase in the first $2^{+}$ excitation energy [$E(2^+_1)$] along the isotopic (isotonic) chain is expected for nuclei with a magic number of neutrons (protons) as compared to their neighbors. Following the same arguments, the reduced electric transition probability $\emph{B}$($\emph{E}$2; 0$_{1}^{+}$$\rightarrow$2$_{1}^{+}$) in even-even nuclei, reflecting how easy it is to excite protons to form the 2$_{1}^{+}$state, exhibits a local minimum for closed-shell nuclei. For rare isotopes, the first experimentally accessible observable to reflect a shell closure is often the $E(2^+_1)$. The first evidence of the $\emph{N}$ = 32 and 34 magic numbers came from the relatively large $E(2^+_1)$ values in \ts{52}Ca and \ts{54}Ca, respectively~\cite{Huck1985,steppenbeck_nature_2013}, as shown in Fig. \ref{fig:e2_charge}(b). The reduced transition probabilities $\emph{B}$($\emph{E}$2; 0$_{1}^{+}$$\rightarrow$2$_{1}^{+}$) could be determined via Coloumb excitations or lifetime measurements. High-resolution $\gamma$-ray detectors are desirable for lifetime determination using techniques such as lineshape analysis or the Recoil Distance Doppler-Shift (RDDS) method, while high-efficiency $\gamma$-ray detectors with good time resolution are crucial for Coloumb excitation measurements with exotic nuclei.

%Closed-shell nuclei are more stable with additional binding energies and are expected to be more compact with a smaller charge radius. 
Two-neutron separation energies ($S_{2n}$) undergo a sudden decrease from the magic to magic-plus-two-neutron nuclei, reflecting the low binding potential experienced by the two valence nucleons. This behavior can be more clearly visible in the variation of empirical neutron shell gaps, defined as $\Delta_{2n} = S_{2n}(N, Z)-S_{2n}(N+2, Z)$, along the isotopic chain. As shown in Fig.~\ref{fig:e2_charge}(c), mass measurements of \ts{51-56}Ca~\cite{Gallant2012, wienholtz_nature_2013, michimasa_prl_2018} have revealed the large $\Delta_{2n}$ across $\emph{N}$ = 32 and 34, respectively, in line with the expected values for shell closures. 

Charge radii often exhibit a local minimum or inflection at magic numbers owing to the tightly-bound nucleons and spherical nature of magic nuclei, as seen for \ts{48}Ca in Fig.~\ref{fig:e2_charge}(d). However, recent laser spectroscopy along the Ca and K isotopes revealed a continuous increase of the charge radii from $\emph{N}$ = 28 to $\emph{N}$ = 32 and 33, respectively, interpreted as challenging the magicity of $\emph{N}$ = 32~\cite{Koszorus2021}, although the measured charge radii of Ca and Ni isotopes showed that the kink in charge radii does not directly reflect the strength of a shell closure \cite{Felix2022}. The challenge to $\emph{N}$ = 32 magicity from the charge radii measurement brought into question in Ref.~\cite{nowacki_ppnp_2021} invokes arguments of proton and neutron radii equalization through the filling of halo-like $p$ orbitals as discussed in~\cite{Bonnard_2016_PRL}. The measured interaction cross sections for $^{42-51}$Ca at the RIBF provide additional evidence for the interplay between proton and neutron radii, suggesting that the excess neutrons in Ca isotopes beyond $\emph{N}$ = 28 stimulate the $^{48}$Ca to swell \cite{Tanaka2020PRL}. 
% FB deleted below, seems a bit too out of focus? 
%The connection between charge radii evolution and the strength of neutron shell closure has been questioned since a surprisingly similar trend of charge radii across the $\emph{N}$ = 28 shell closure in Ca and Ni isotopes was observed even though \ts{56}Ni is supposed to be much softer than \ts{48}Ca.

Another important measure of magicity is the occupancy of single-particle orbitals. In the independent particle model (IPM), for doubly-magic nuclei, orbitals lying below the shell closure are fully occupied with 2$J$+1 nucleons. Spectroscopic factors, which are a measure of the wavefunction overlap between two neighboring nuclei, can be inferred through one-nucleon removal reactions from doubly-magic nuclei. The results of these measurements reflect the single-particle occupancy, which in turn informs on the robustness of the shell closure in question. In Fig.~\ref{fig:sf}, the ratio between the measured $(p,pn)$ cross-section and the calculated single-particle cross-section normalized to (2$J$+1) are used to study the magicity of $^{48,52,54}$Ca. 

% Deleted by FB, as its repeated later on word for word
%These three Ca isotopes show a very similar pattern: a ratio close to unity below the Fermi level and a very small ratio above, suggesting that the $\emph{N}$ = 32 shell closure in \ts{52}Ca and the $\emph{N}$ = 34 shell closure in \ts{54}Ca is as strong as $\emph{N}$ = 28 in \ts{48}Ca.
%Shell closure Of manifestations of shell evolution the most striking is the formation of substantial shell closures which lead to step changes of otherwise systematic behaviour. Generally speaking, the first experimentally accessible observable to reflect a shell closure is the $E(2^+_1)$, which follows a smooth trend across semi-magic isotonic and isotopic chains, but is suddenly enhanced when reaching a doubly magic nucleus.  
\section{Theoretical calculations}

Several theoretical models have been employed to interpret the experimental results related to the $\emph{N}$ = 32, 34 shell imigration within the context of direct reactions at the RIBF with liquid hydrogen targets. Broadly speaking, two families of nuclear structure calculations were used, including the shell model with configuration interactions and the {\em ab initio} approaches based on the realistic nuclear forces. For the description of the quasi-free scattering reaction process, the distorted-wave impulse approximation (DWIA), discussed in detail in another dedicated article \cite{yoshida-ptep} of this special issue, was adopted. 

%Several theoretical models have been employed to interpret the experimental results related to the 
%$\emph{N}$ = 32, 34 shell closures in direct reactions at the RIBF with liquid hydrogen targets. Broadly speaking, two families of nuclear structure calculations shall be considered: the effective shell model and {\em ab initio}. The distorted-wave impulse approximation, used for the reaction mechanism, shall be discussed separately in detail within this series of articles

% shell model 
The possible existence of $\emph{N}$ = 32, 34 subshell closures was initially suggested by nuclear structure calculations using the effective shell-model interactions. In particular, the GXPF1 family of interactions plays an important role in motivating the search for the novel $\emph{N}$ = 34 magic number, with its prediction quantitatively suggested in its GXPF1A iteration~\cite{honma_epja_2005}, after its qualitative suggestion from the role of tensor forces~\cite{otsuka_prl_2001}. In the following discussions relevant to neutron-rich Ca isotopes, the GXPF1Br~\cite{steppenbeck_nature_2013}, GXPF1Bs~\cite{chen_prl_2019} and A3DA-m~\cite{tsunoda_prc_2014} interactions were used, while for isotopes with $\emph{Z}$ < 20, the SPDF-MU~\cite{utsuno_prc_2012}, SPDF-MUr~\cite{utsuno2014} and SDPF-U~\cite{nowacki2009} interactions were used. All these effective interactions originate within the microscopic G-matrix theory~\cite{hjorthjensen_pr_1995} but in different model spaces and with different empirical corrections to the matrix elements and single-particle energies.

 %The role of the tensor interaction is significant in the description of the $\emph{N}$ = 32, 34 shell closures. %It is then important that calculations can incorporate it explicitly. Shell-model Hamiltonians are able to incorporate effectively this from microscopic models into effective interactions~\cite{honma_rapr_2008}. 

% Ab initio
Approaches that derive interactions from microscopic exchange theories, with minimal input from experiment are here termed as {\em ab initio}. Finding relatively recent applications in nuclear structure, the {\em ab initio} approach mainly employs nucleon-nucleon and three-nucleon interactions derived from chiral effective field theory (EFT)~\cite{epelbaum_rmp_2009,machleidt_pr_2011} that facilitates an efficient expansion and consistent treatment of both nuclear forces and electroweak currents. Whilst the interactions are derived in a similar way, the methods to solve the many-body problem vary significantly among different {\em ab initio} calculations such as Self-Consistent Green's Functions (SCGF)~\cite{barbieri_lp_2017}, coupled cluster (CC)~\cite{Hagen2012PRL}, and in-medium similarity renormalization group (IMSRG)~\cite{holt_prc_2014}. The selection of the solution methods highlights different aspects of nuclear structure elements at play. For instance, SCGF features a direct link with one-nucleon removal reactions and can provide reliable predictions for spectroscopic factors.

%both the aforementioned nuclear structure approaches are utilized to calculate spectroscopic factors. 
In the case of the proton-induced one-nucleon removal reaction, the single-particle cross sections and parallel momentum distributions are calculated in the DWIA framework, which is the most widely used theoretical method to describe the $(p,pN)$ reaction. DWIA assumes a single interaction of the incident proton with the struck nucleon in the reaction process and has been first applied to
$(p,pN)$ reaction in inverse kinematics with the eikonal distorted wave functions(eikonal DWIA) ~\cite{aumann_prc_2013} and later in the standard partial-wave expansion form (DWIA) ~\cite{ogata_prc_2015,wakasa_ppnp_2017}. The eikonal DWIA calculations for $(p,2p)$ have been shown to provide consistent reduction factors, defined as the ratios of experimental to theoretical one-nucleon removal cross sections, as extracted from $(e,e’p)$ measurements~\cite{atar2018}. However, these calculations under the adiabatic and eikonal approximations always yield symmetric parallel momentum distributions. Conversely, DWIA calculations that take into account nonlocality and relativistic effects can describe the observed asymmetric parallel momentum distributions from $(p,pN)$ reactions at lower incident beam energies~\cite{pohl2023}, but DWIA gives larger reduction factors than those reported by $(e,e’p)$ analysis~\cite{phuc2019}. Due to systematic discrepancies among reaction models \cite{AUMANN2021}, it is essential to discuss the relative change of spectroscopic factors within a consistent theoretical framework.\\

\section{Experimental results for shell migration at $\it{N}$ = 32, 34 and beyond}

We now present an overview of the experimental progress in the study of shell migration at $\emph{N}$ = 32, 34, and towards $\emph{N}$ = 40 for neutron-rich $pf$-shell nuclei, achieved from quasi-free scattering measurements using the LH$_2$ target. The radioactive nuclei of interest were produced by fragmentation of a 345 MeV/u $^{70}$Zn primary beam with an average beam intensity of 240 pnA on a 10-mm-thick rotating Be target. These selected results aim to highlight the benefits of the application of direct reactions on a non-composite target. 

\subsection{Shell closure at $\it{N}$ = 32}

\begin{figure}[bth]
\centering
\includegraphics[width=1.0\textwidth]{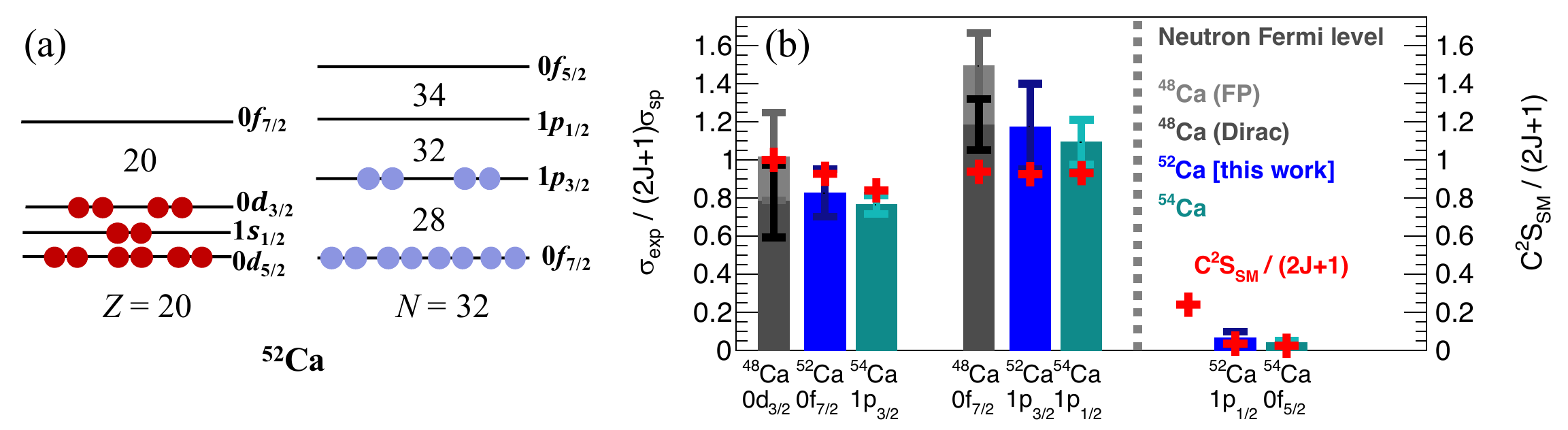}
    \caption{ (a) Configurations of valence nucleons in $^{52}$Ca. (b) The ratio of the measured $^{48,52,54}$Ca$(p,pn)$ cross-section and the calculated single-particle cross-section normalized to (2J+1) below and above the corresponding shell closures. Reprinted figure from \cite{Enciu2022}. Copyright CC BY 4.0.}%Reprinted figure with permission from \cite{Enciu2022}. Copyright 2022 by the American Physical Society.}
    \label{fig:sf}
\end{figure}

%\begin{itemize}
%    \item $N=32$ shell closure (Experimental): 
%    \begin{itemize}
%        \item Re-measured by Ref.~\cite{Perrot2006}, also observed 2.22~MeV state in \textsuperscript{53}Ca, $\beta$-decay
%        \item First in-beam by Ref.~\cite{Gade2006}, cross-shell excitation, 2p knockout, "confirmed" doubly-magic nature
%        \item Deep inelastic scattering by Ref~\cite{Rejmund2007}, based on $1p_{1/2}-0f_{5/2}$ systematics suggested no $N=34$
%        \item Mass measurement at ISOLDE \cite{wienholtz_nature_2013}, $S_{2n}\Rightarrow$ good $N32$
%        \item Mass measurements at TRIUMF \cite{leistenschneider_prl_2021} imply abrupt onset of $N=32$ from Ti to Sc. 
%        \item Laser spectroscopy \cite{Koszorus2021}, charge radius implies no $N=32$ in K
%        \item \textbf{Madalina uses knockout to show extended orbital but good shell closure \cite{Enciu2022}}
%    \end{itemize}
%A major difference between these approaches is the latter's assumption of the adiabatic approximation, leading to necessarily symmetric parallel momentum distributions. On the other hand, the DWIA incorporates non-adiabaticity, which is then able to describe the observed asymmetric parallel momentum distributions. Within the context of the present article, this distinction is important in the assignment of spin-parities to excited and ground states populated through knockout reactions. 
%\section{Shell migration at N = 32, 34 and towards N = 40}
% FB changed this as it's the title of the whole contribution
The $\emph{N}$ = 32 subshell closure is a direct consequence of the weakening attractive ${\pi}0f_{7/2}$-${\nu}0f_{5/2}$ interactions, but still originate from the spin-orbit 
splitting between 1$p_{3/2}$ and 1$p_{1/2}$ neutron orbitals, as shown in Fig.~\ref{fig:sf}. The measured systematics of the $E$(2$_{1}^{+}$)~\cite{Steppenbeck2015PRL, Huck1985, Kanungo2002PLB, Janssens2002PLB, Prisciandaro2001PLB, cortes2020prc} along the isotopic chain exhibits a local maximum at $^{52}$Ca, $^{54}$Ti, and $^{56}$Cr, as shown in Fig.\ref{fig:e2_charge}(b), supporting the existence of the $\emph{N}$ = 32 sub-shell closure in these isotopes. The measured reduced transition probabilities~\cite {Dinca2005PRC, Burger2005PLB} in Ti and Cr isotopes reveal a local minimum at $\emph{N}$~=~32, in line with their 2$^{+}$ results. With the development of mass measurement techniques like Penning trap and multi-reflection time-of-flight, the mass spectroscopy of neutron-rich K, Ca, Sc, Ti, V and Cr isotopes has been achieved ~\cite{wienholtz_nature_2013,Rosenbusch2015PRL,Xu_2015_CPC,Leistenschneider2018PRL,Xu2019,Leistenschneider2021,Iimura2023}. As shown in Fig.\ref{fig:e2_charge}(c), the extracted empirical neutron shell gap $\Delta_{2n}$ suggests that the $\emph{N}$ = 32 subshell closure peaks at $^{52}$Ca due to its doubly magic nature, is still strong in $^{51}$K and $^{53}$Sc, gets weaker in $^{54}$Ti, and gradually vanishes for V and Cr isotopes.

As mentioned in Section 2, the charge radii measurement for neutron-rich Ca and K isotopes shows no kink at $\emph{N}$ = 32  ~\cite{GarciaRuiz2016NaturePhy,Kreim2014PLB,Koszorus2021}, interpreted as challenging the $\emph{N}$ = 32 magicity. Bonnard and his collaborators proposed a possible explanation~\cite{Bonnard_2016_PRL}. An effective increase in size of $1p_{3/2}$ neutron orbital, in turn influencing the proton radial extention due to the isovector polarizability, can reproduce the observed charge radii of K and Ca isotopes, while maintaining the $\emph{N}$ = 32 subshell closure. A sizeable difference of $\sim$0.7\,fm was predicted between the root-mean-square (rms) radius of the $1p_{3/2}$ and the $0f_{7/2}$ neutron orbitals. In quantum mechanics, position and momentum operators do not commute and follow the Heisenberg uncertainty principle. Momentum is indeed just the Fourier transform variable of position. In $(p,pN)$ reaction, the momentum distribution of the fragment is equivalent to that of the removed nucleon. It relates to the single-particle wave function and, consequently, to the root mean square radii of the knocked-out nucleon.

\begin{figure}[bth]
\centering
\includegraphics[width=0.8\textwidth]{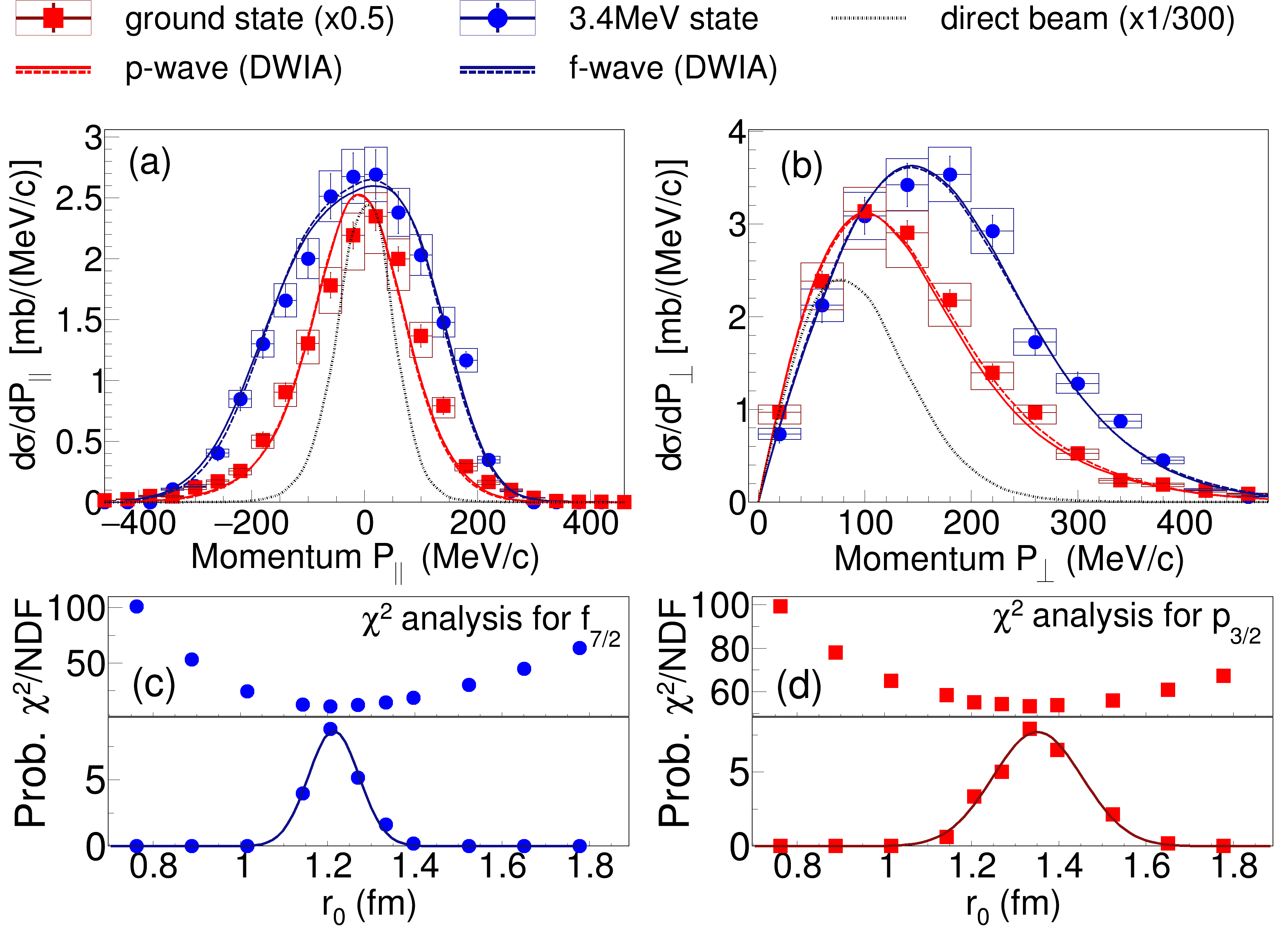}
    \caption{ Experimental parallel (a) and perpendicular (b) momentum distributions of the $^{52}$Ca direct beam (dotted black line), $^{51}$Ca ground-state (red squares), and 3453-keV state (blue circles) population together with the theoretical curves for $p$-wave (red) and $f$-wave (blue), with a binning of 40 MeV/c. The calculations were performed using a folding potential (solid lines) and the Dirac phenomenology potential (dashed lines) with $a_0=0.67$\,fm and the optimal $r_0$ values: 1.35\,fm ($p$-wave) and 1.21\,fm ($f$-wave). The statistical errors are marked with crosses and the systematic errors on the absolute normalization with boxes. The (c) and (d) panels show the reduced $\chi^2$ (upper panels), i.e., $\chi^2$/NDF (NDF being the number of degrees of freedom), and the probability distribution (lower panels) for the $0f_{7/2}$ and $1p_{1/2}$ orbitals as a function of the parameter $r_{0}$. Reprinted figure from \cite{Enciu2022}. Copyright CC BY 4.0.}%Reprinted figure with permission from \cite{Enciu2022}. Copyright 2022 by the American Physical Society.}
    \label{fig:PMD_52ca}
\end{figure}

To extract the spatial distribution of the valence neutrons in $^{52}$Ca and investigate the robustness of its double magicity, the quasi-free one-neutron knock-out $^{52}$Ca$(p,pn)$$^{51}$Ca on the MINOS liquid hydrogen was performed at $\sim$\,230\,MeV/nucleon~\cite{Enciu2022}. The average intensity of the $^{52}$Ca beam was 4.4 particles per second (pps). The measured coincident $\gamma$ rays using DALI2+ array revealed a prominent $\gamma$-ray transition at 3453(20)\,keV together with a few weak transitions at energies of 2375(13), 1720(25), 1461(20), and 691(4)\,keV, consistent with previous high-resolution decay-spectroscopy measurements~\cite{fornal_prc_2008}. After subtracting the neutron-evaporation contribution from the inelastic scattering process, the partial $^{52}$Ca$(p,pn)$ cross sections to individual states were obtained by fitting the measured $\gamma$-ray intensities. The strong population of the ground and 3453-keV states is consistent with neutron knockout from the $1p_{3/2}$ and $0f_{7/2}$ orbitals, respectively, while the weak population of the 1720-keV state corresponds to the neutron knockout from the $1p_{1/2}$ orbital. The ratio between the measured exclusive quasi-free cross-section and the calculated single-particle cross-section normalized to (2$J$+1) were used to assess the occupancies of valence neutrons in the $1p_{3/2}$, $0f_{7/2}$ and $1p_{1/2}$ orbitals in \ts{52}Ca. For a systematic comparison, the \ts{48,54}Ca$(p,pn)$ data from Refs.~\cite{chen_prl_2019,ahmad_npa_1984} were also analyzed using the same DWIA formalism. The results shown in Fig.~\ref{fig:sf} reveal a very similar pattern for these three Ca isotopes: a ratio close to unity below the Femi level and a very small ratio above, suggesting that the $\emph{N}$ = 32 shell closure in $^{52}$Ca is as strong as the $\emph{N}$ = 34 shell closure in $^{54}$Ca and $\emph{N}$ = 28 in $^{48}$Ca.

%The momentum distribution (MD) of the knocked out nucleon is linked with its wave function's spatial extension when it is still bound to the pre-reaction nucleu, therefore the root-mean-square (rms) radii via the Fourier transformation based on the uncertainty principle. 
The parallel and perpendicular momentum distributions (PMDs) of the $^{51}$Ca fragments from the $(p,pn)$ reaction to the ground state ($1p_{3/2}$) and the 3.4-MeV state ($0f_{7/2}$) were constructed by gating on each bin of the inclusive momentum and fitting the corresponding $\gamma$-ray spectrum and shown in Fig.~\ref{fig:PMD_52ca}. Theoretical momentum distributions were calculated within the DWIA framework as introduced in Section 3, and convoluted with the experimental momentum resolution (49.5\,MeV/$c$ for parallel and 76.9\,MeV/$c$ for perpendicular). PMDs were found sensitive to the $r_0$ term of the Woods-Saxon potential used for the calculation of the single-particle wave function of the knocked-out neutron, while not sensitive to $a_0$. To extract the rms radius of the $1p_{3/2}$ and $0f_{7/2}$ orbital, a $\chi^2$ criterion was used and a probability analysis assuming a Gaussian probability density function was performed by fitting the experimental PMDs with theoretical curves calculated using different $r_0$ values. Figures ~\ref{fig:PMD_52ca}(c) and (d) show the results with a clear dependence on $r_0$. The deduced optimal $r_0$ and associated 1-$\sigma$ uncertainty gives an rms radii of 4.74(18)\,fm and 4.13(14)\,fm for the single-particle wave function of the knocked-out neutron for $1p_{3/2}$ and $0f_{7/2}$, respectively. Therefore, the analysis of the momentum distributions gives a difference of the rms radii of the neutron $0f_{7/2}$ and $1p_{3/2}$ orbitals of 0.61(23)\,fm, in agreement with the modified shell-model prediction of 0.7\,fm suggesting that the large rms radius of the $1p_{3/2}$ orbital in neutron-rich Ca isotopes is responsible for the unexpected linear increase of the charge radius with the neutron number~\cite{Bonnard_2016_PRL}. 

In addition, the proton configurations in $^{52}$Ca were studied using the $^{52}$Ca$(p,2p)$$^{51}$K reaction with the MINOS liquid hydrogen at $\sim$\,250\,MeV/nucleon at the RIBF~\cite{Sun2020PLB}. The same DWIA calculations as used for $^{52}$Ca$(p,pn)$$^{51}$Ca were adopted. The measured exclusive parallel momentum distributions enabled unambiguous spin-parity assignment for the ground state (3/2$^{+}$) and the first excited state (1/2$^{+}$) in $^{51}$K. The results confirmed the restoration of the natural ordering of the 1/2$^{+}$ and 3/2$^{+}$ proton hole state at $\emph{N}$ = 32, different from the inversion observed at $\emph{N}$ = 30. The ratio between the measured partial cross section and the calculated single-particle cross section normalized to (2$J$+1) was found to be 0.75(6) and 0.45(5) for the 3/2$^{+}$ and 1/2$^{+}$ states, respectively, in line with the closed $\emph{Z}$ = 20 shell configurations. 

Below ${Z}$ = 20 in Ar isotopes, the persistence of the $N=32$ subshell closure was first indicated in Ref.~\cite{Steppenbeck2015PRL} through the measurement of the $E(2^+_1)$ of $^{50}$Ar. However, no peak was observed at $\emph{N}$ = 32, as shown in Fig.\ref{fig:e2_charge}(c). The result was interpreted as arising from the strong mixing of different configurations in the wave function of the 0$_{1}^{+}$ and 2$_{1}^{+}$ state in $^{50}$Ar based on SDPF-MU calculations, although its $\emph{N}$ = 32 gap is of similar magnitude to that of $^{52}$Ca. During the work of reactions with the MINOS LH\tb{2} target at the RIBF, the level scheme of $^{50}$Ar was expanded to seven levels through single- and multi-nucleon knockout reactions~\cite{cortes2020prc}. The majority of these states were tentatively assigned as $2^+$, with a candidate $3^-$ state determined through the inelastic scattering channel. The spin-parity assignments were based on a comprehensive comparison between measured results, including energies and populated cross sections from the direct $(p,2p)$ and $(p,pn)$ reactions, with theoretical values from SPDF-MU and VS-IMSRG, folded with DWIA calculations~\cite{cortes2020prc}. The consistency emphasizes the emergency of the $\emph{N}$ = 32 subshell closure in $^{50}$Ar.

\subsection{Shell closure at $\it{N}$ = 34}

%   \item $N=34$ shell closure: 
%    \begin{itemize}
%        \item Predicted by \cite{otsuka_prl_2001} from features of tensor
%        \item High $E(2^+_1)$ by \cite{steppenbeck_nature_2013} indicated shell closure, also $3^-$ implied. 12 pps of \textsuperscript{55}Sc, and 125 pps of \textsuperscript{56}Ti $\Rightarrow$ 2p knockout
%        \item Mass measurements \textsuperscript{55-57}Ca, $S_{2n}\Rightarrow$ good $N=34$ \cite{michimasa_prl_2018}
%        \item \textbf{Chen uses \textsuperscript{54}Ca(p,pn) to show locality of $1p_{3/2}\Rightarrow$ good $N=34$ \cite{chen_prl_2019}}
%        \item \textbf{Hongna measures \textsuperscript{52}Ar $E(2^+_1)\Rightarrow$ implies $N=34$ in $Z<20$ \cite{liu_prl_2019}}
%        \item \textbf{Frank measures \textsuperscript{55}Sc(p,2p) and \textsuperscript{55}Ca(p,pn), feedback on structure of \textsuperscript{54}Ca energy levels \cite{browne_prl_2021}}
%    \end{itemize}
%\end{itemize}
Unlike the $\emph{N}$ = 32 subshell closure, which can be understood through spin-orbit splitting, the $\emph{N}$ = 34 energy gap has been interpreted through effective shell-model interactions as arising from the absence of tensor-force attraction between $\pi0f_{7/2}$ and $\nu0f_{5/2}$ orbitals~\cite{otsuka_prl_2001}.
The $\emph{N}$ = 34 subshell closure had only been suggested as present in $^{54}$Ca~\cite{steppenbeck_nature_2013,michimasa_prl_2018}, before the study of \ts{52}Ar presented below.
In the Ti and Cr isotopes, the measured systematics of $\emph{E}$(2$^{+}_{1}$)~\cite{Suzuki2013PRC,Zhu2006PRC} and $\emph{B}$($\emph{E}$2; 0$_{1}^{+}$$\rightarrow$2$_{1}^{+}$) \cite{Dinca2005PRC, Burger2005PLB} show no local maximum and minimum at $\emph{N}$ = 34. The mass measurements of Sc, Ti, Cr and V isotopes do not support the existence of the $\emph{N}$ = 34 subshell closure in $\emph{pf}$-shell nuclei at atomic number $\emph{Z}$ $>$ 20, in line with the measured low-lying structure of $^{55}$Sc~\cite{Steppenbeck2017PRC} indicating a rapid weakening of the $\emph{N}$ = 34 subshell closure above $\emph{Z}$ $=$ 20. The $\emph{E}$(2$_{1}^{+}$) of $^{54}$Ca was measured to be 2043(19)\,keV, $\sim$0.5\,MeV lower than $^{52}$Ca~\cite{steppenbeck_nature_2013}, but much higher than \ts{56}Ti. Despite this lower 2$^{+}_{1}$ excitation energy, $^{54}$Ca was concluded to be a doubly magic nucleus from a phenomenological shell-model interpretation~\cite{steppenbeck_nature_2013}, whereas $\emph{ab initio}$ coupled-cluster calculations indicated a weak $\emph{N}$~=~34 subshell closure~\cite{Hagen2012PRL}. Very recently, the mass measurements of $^{55-57}$Ca~\cite{michimasa_prl_2018} confirmed the picture of a subshell closure at $\emph{N}$ = 34 in Ca isotopes, and suggesting the energy gap between the neutron $1p_{1/2}$ and $0f_{5/2}$ orbitals in $^{54}$Ca being comparable to the gap between the neutron $1p_{3/2}$ and $1p_{1/2}$ orbitals in $^{52}$Ca. It is thus interesting to study how strong the $\emph{N}$ = 34 subshell closure in \ts{54}Ca and how the $\emph{N}$ = 34 subshell evolves below $\emph{Z}$ = 20 towards more neutron-rich systems, such as \ts{52}Ar.

%%%% 54Ca(p,pn)53Ca
\begin{figure}[bt]
\centering
\includegraphics[width=0.99\textwidth]{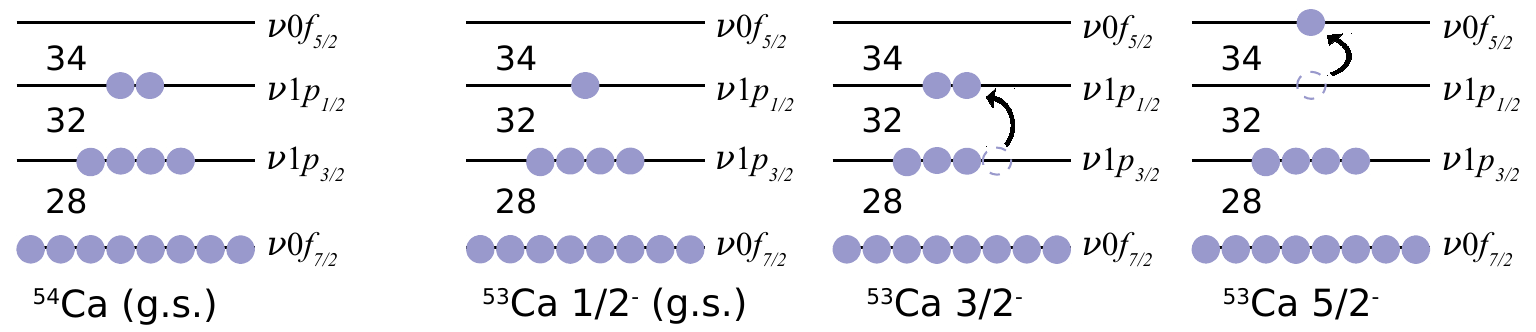}
    \caption{Illustration of the most representative neutron single-particle configurations for ground states of $^{54}$Ca, and ground and bound excited states of $^{53}$Ca. Reprinted figure with permission from \cite{chen_prl_2019}. Copyright 2019 by the American Physical Society.}
    \label{fig:54Ca53Caconfig}
\end{figure}

\begin{table}[tb]
\centering
\caption{\label{tab:cx}Inclusive and exclusive cross sections (in mbarn) for the
\ts{54}Ca($p$,$pn$)\ts{53}Ca reaction ($\sigma$\tb{-1$n$}),
compared with theoretical values ($\sigma_{\textrm{-1}n}^\textrm{th}$)
using the calculated single-particle cross sections ($\sigma$\tb{sp}) from the DWIA framework
and spectroscopic factors ($C^{2}S$) from SM and $\emph{ab initio}$ calculations. Reprinted table with permission from \cite{chen_prl_2019}. Copyright 2019 by the American Physical Society.}
\begin{tabularx}{\textwidth}{lccccXXXXXX}
\hline
&  &  &  & DWIA  & 
\multicolumn{2}{c}{GXPF1Bs}   & 
\multicolumn{2}{c}{NNLO\tb{sat}} &
\multicolumn{2}{c}{$N\!N$+3$N$(lnl)} \\ 
\cline{6-7} \cline{8-9} \cline{10-11}
& $J^{\pi}$ & -1$n$ & $\sigma$\tb{-1$n$} & $\sigma_\textrm{sp}$ 
& $C^{2}S$ & $\sigma_{\textrm{-1}n}^\textrm{th}$ 
& $C^{2}S$ & $\sigma_{\textrm{-1}n}^\textrm{th}$ 
& $C^{2}S$ & $\sigma_{\textrm{-1}n}^\textrm{th}$ \\
\hline
g.s.     & $1/2^-$ & $1p_{1/2}$ & 15.9(17) & 7.27 &
1.82 & 13.2 & 
1.56 & 11.3 &
1.58 & 11.6 \\
2220(13) & $3/2^-$ & $1p_{3/2}$ & 19.1(12) & 6.24 &
3.55 & 22.2 & 
3.12 & 18.5 &
3.17 & 17.0 \\
1738(17) & $5/2^-$ & $0f_{5/2}$ &\phantom{0}1.0(3)\phantom{0} & 4.19 & 
0.19 & \phantom{0}0.8 & 
0.01 & \phantom{0}0.1 &
0.02 & \phantom{0}0.1 \\
Inclusive &  &  & 36.0(12) &  &
& 36.2 & 
& 29.9 &
& 28.7 \\
\hline
\end{tabularx}
\end{table}

The ground state properties of $^{54}$Ca were studied by performing the quasi-free one-neutron knockout reaction $^{54}$Ca$(p,pn)$$^{53}$Ca using the MINOS liquid hydrogen target~\cite{chen_prl_2019}. The $^{54}$Ca particles were produced with a mean intensity of 7.3 pps.
The most representative neutron configurations for the ground states of $^{54}$Ca, and the ground and excited states of $^{53}$Ca are shown in Fig.~\ref{fig:54Ca53Caconfig}.
In a simple single particle shell model picture, the ground state of $^{53}$Ca has the unpaired neutron occupying the ${\nu}1p_{1/2}$ orbital ($1/2^{-}$), and the two excited states observed in previous experiments~\cite{steppenbeck_nature_2013,Perrot2006} have the configurations of one hole in the ${\nu}1p_{3/2}$ orbital ($3/2^-$) and the unpaired neutron occupying the ${\nu}0f_{5/2}$ orbital ($5/2^-$).
By doing one-neutron knockout from the ground state of $^{54}$Ca ($0^+$), the population of each $^{53}$Ca state corresponds to the neutron removal from one specific orbital.
In this experiment, the populated $^{53}$Ca states were tagged by the in-beam $\gamma$-ray measurements using the DALI2+ array. 
The momentum distributions of the $^{53}$Ca residues in the center of mass frame of $^{54}$Ca were extracted for individual $^{53}$Ca states, and compared with the theoretical calculations from DWIA to distinguish the neutron removal orbitals.
The exclusive cross sections to individual $^{53}$Ca states were extracted to probe the ground state wave function of $^{54}$Ca.
The measured cross section to the $1p_{3/2}$ state of $^{53}$Ca is about 20 times larger than the one to the $0f_{5/2}$ state, suggesting a small $f$-wave component in the ground state of $^{54}$Ca, therefore, providing direct experimental evidence of a strong $\emph{N}$ = 34 subshell closure.
The measured cross sections were well reproduced by the theoretical calculations from DWIA reaction model with structure inputs from the shell model using GXPF1Bs interaction and $\emph{ab initio}$ calculations using NNLO\tb{sat} and $N\!N$+3$N$(lnl) interactions, as shown in Table~\ref{tab:cx}.

With the firm arguments detailed above of a doubly-magic \ts{54}Ca, its detailed structure is of interest to further benchmark the models that predict the $N=34$ shell closure. Investigations of the structures of excited states of \ts{54}Ca were carried out using the direct reactions: \ts{55}Sc($p$,2$p$)~\cite{browne_prl_2021} and \ts{55}Ca($p$,$pn$)~\cite{browne_2024_55Ca54Ca}. Whilst the former reaction populated negative parity states built from proton-hole configurations, the latter populated positive-parity neutron-hole states. The structure of the populated states were identified, where possible, through the parallel momentum distributions of the outgoing products. In the case of the reaction populating a bound state, the parallel momentum of the \ts{54}Ca was measured, and in the case of neutron-unbound states the combined momentum of the residual \ts{53}Ca+$n$ system was considered. 

%\begin{figure}[bt]
%\centering
%\includegraphics[width=0.49\textwidth]{Figs/xs_p2p.pdf}
%\includegraphics[width=0.49\textwidth]{Figs/xs_ppn.pdf}
    %\caption{Left: Experimental and theoretical results from the \ts{55}Sc(p,2p) reaction. Right: Same as left for \ts{55}Ca(p,pn) reaction. (Top panels) Below $S_{n}$, the measured cross sections to states from $\gamma-ray$ spectroscopy are shown. Above $S_{n}$, the population cross sections of the unbound strengths from invariant mass spectroscopy are shown at the energy centroids of fitted values and likely represent contributions from several states. States with conclusive PMD are colored accordingly to the $\ell$-value of their removed nucleon, otherwise are black. (Bottom panel) Theoretical predictions of state energies and their population cross sections are shown.}
%    \label{fig:54ca_xs}
%\end{figure}

\begin{figure}[bt]
\centering
\includegraphics[width=\textwidth]{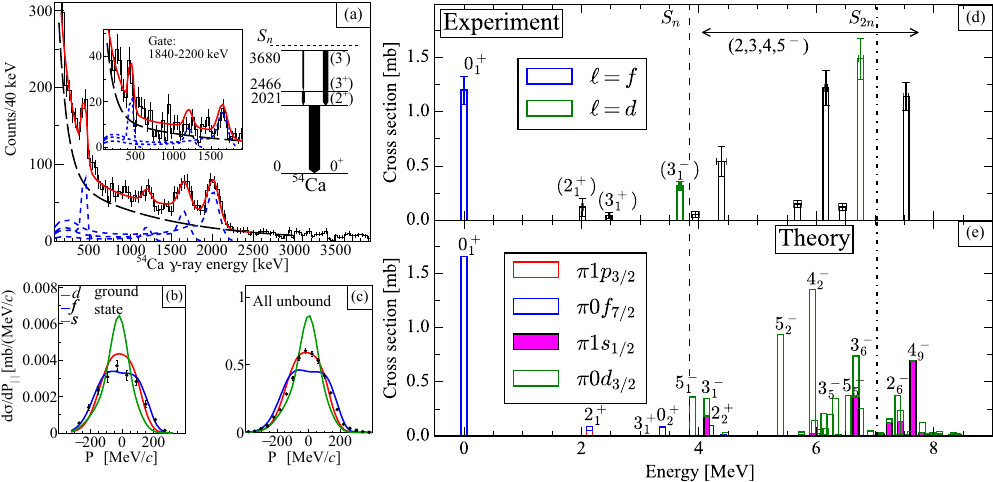}
    \caption{(a) $\gamma$-ray energy spectrum measured in coincidence with the \ts{55}Sc($p$,2$p$)\ts{54}Ca reaction. Inset are the coincidence spectrum observed by gating on the 2021-keV transition and the deduced level scheme. (b) The parallel momentum distribution measured in coincidence with the population of the ground state of \ts{54}Ca. The points are experimentally derived, whilst the lines are from DWIA calculations assuming different orbital removals. (c) Same as (a) but for the sum of the unbound states. (d) Observed cross sections to states populated by the \ts{55}Sc($p$,2$p$) reaction. Where a parallel momentum distribution could be extracted for a state, the colour indicates the orbital removed to populate the state. (e) Predicted cross sections to states in \ts{54}Ca from the \ts{55}Sc($p$,2$p$) reaction. The cross sections are the product of the spectroscopic factors from the GXPF1Br shell model calculations and the single-particle nucleon removal cross sections from the DWIA. Reprinted figure with permission from Ref.~\cite{browne_prl_2021}. Copyright 2021 by the American Physical Society.}
    \label{fig:54ca_xs}
\end{figure}

Figure~\ref{fig:54ca_xs}(a) shows the $\gamma$-ray energy spectrum measured in coincidence with the \ts{55}Sc$(p,2p)$ reaction. The results agree with and expand, though the definite measurement of the 445~keV transition, the level scheme measured in Ref.~\cite{steppenbeck_nature_2013}. The neutron decays of unbound states were also measured, the relative energy spectrum can be found in Ref.~\cite{browne_prl_2021}. Parallel momentum distributions of the ground state and the excited unbound states are shown in Figs.~\ref{fig:54ca_xs}(b) and (c), respectively. It is evident that the ground state of \ts{54}Ca is populated following the removal of the valence $0f_{7/2}$ proton, and excited states populate from the removal of $0d_{3/2}$. It is worth noting that the parallel momentum distribution associated with the population of the $(3^-)$ state, shown in Ref.~\cite{browne_prl_2021}, follows the removal of a $d$-wave proton. The measured energy levels and their population cross sections compared to the theoretically-derived values are shown in Fig.~\ref{fig:54ca_xs}(d) and (e). The calculated energies are from shell model calculations employing the GXPF1Br interaction and the cross sections as the product of single-particle cross sections from DWIA and spectroscopic factors from the shell model. The calculations reflect very well the observed energies and cross sections of population. 

The results from the \ts{55}Sc$(p,2p)$ reaction show that the removal of the valence $0f_{7/2}$ proton populates only the ground state of \ts{54}Ca, despite the ground state of \ts{55}Sc containing significant excited-state configurations. Moreover, the data suggests that all populated excited-states are from the removal of a $\pi0d_{3/2}$ orbital. %The provides the first experimental evidence of the $J^\pi=3^-$ nature of the state at $\sim3.7$~MeV, as well as clearly identifying the $\sim450$~keV transition only alluded to in Ref.~\cite{steppenbeck_nature_2013}. %Moreover, it showed a rather dense level structure beyond $S_n$, reflecting the predicted fragmentation strength of the $\pi0d_{3/2}$ hole state. 

%The \ts{55}Ca(p,pn) data provides further evidence of the $\nu0f_{5/2}$ orbital being the valence particle of \ts{55}Ca, and a very clear demonstration of the existence of the $\sim450$~keV transition and its coincidence with the $2^+_1\rightarrow0^+_\text{g.s.}$ transition. The PMD shows clearly that the state at $\sim2450$~keV is populated from the removal of a $p$-wave orbital with the structure of $(\nu0f_{5/2})^1(\nu1p_{1/2})^{-1}$, where the spins are aligned, implying the $2^+_1$ state has the same configuration with an anti-aligned spin coupling. Between $S_n$ and $S_{2n}$ lies the dominant population strength of states whose PMD shows again a clear $p$-wave structure, and owing to the exhaustion of spin-configurations for the bound states is assigned to removal of the $\nu0p_{3/2}$ orbital. The distribution of populated states between $S_n$ and $S_{2n}$ agrees remarkably with the calculated values. Finally, the broad level strength whose centroid is observed beyond $S_{2n}$, despite having low statistics, could be assigned to $f$-wave neutron removal owing to its isolation in the $E_\text{rel}$ spectrum. Comparison to the calculations show this strength aligns well with the predicted multiplet of states from the $(\nu0f_{5/2})^1(\nu0f_{7/2})^{-1}$ configuration. %The experimental results of the \ts{55}Ca(p,pn) reaction provide strong evidence of the composition of the wavefunction of the excited neutron states in 

\begin{figure}[bt]
\centering
\includegraphics[width=0.98\textwidth]{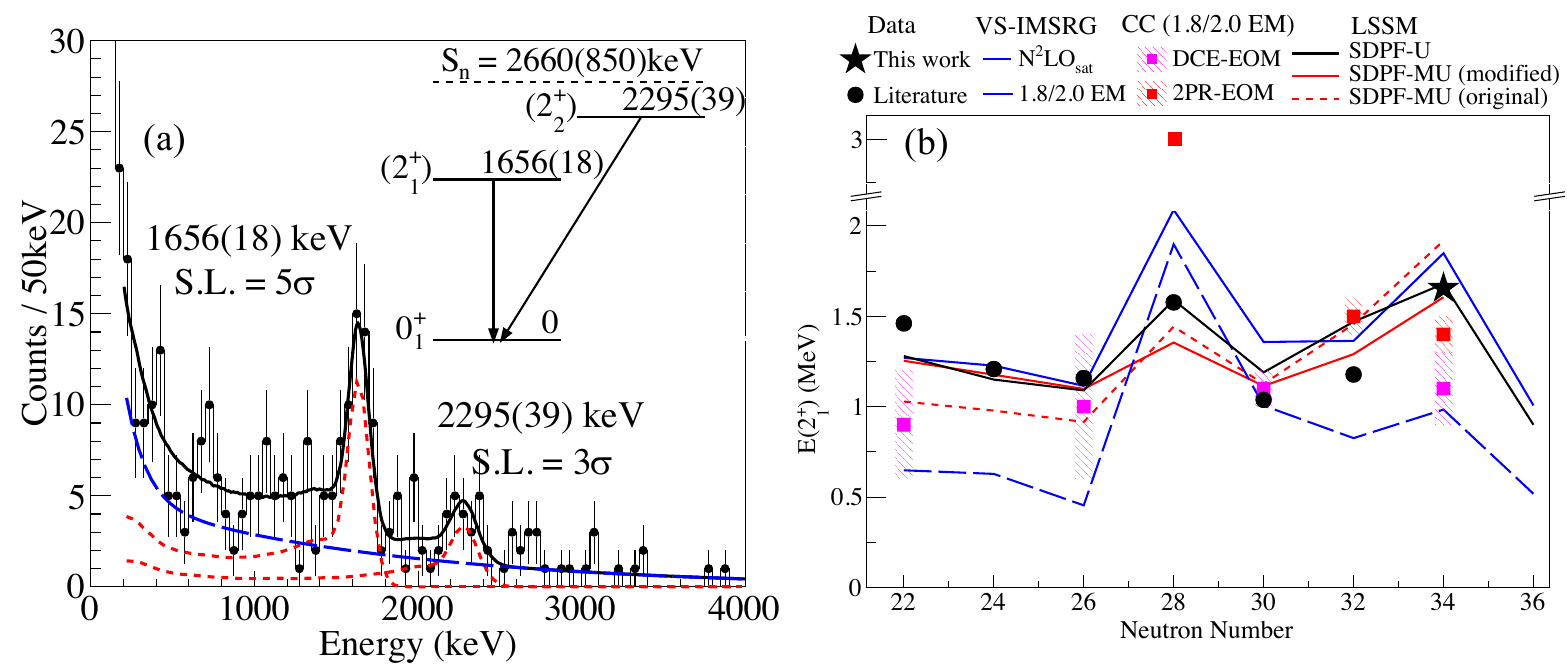}
    \caption{(a) The measured $\gamma$-ray energy spectrum of $^{52}$Ar from the $^{53}$K($\emph{p}$,2$\emph{p}$) reaction. The significance level (S.L.) is given for the observed transitions. (b) Comparison between measured and calculated $\emph{E}$(2$^{+}_{1}$) for even-even Ar isotopes. Calculations are done using VS-IMSRG with the 1.8/2.0 (EM)and N$^{2}$LO$_{sat}$ interactions, coupled-cluster calculations (CC) using the DCE-EOM and 2PR-EOM methods with the 1.8/2.0(EM) interaction, and large-scale shell model (LSSM) calculations with the SDPF-U and the so-called original and modified SDPF-MU interactions. The hatched regions represent the theoretical uncertainties in coupled-cluster calculations. Note the broken y-axis scale between 2.2 and 2.7 MeV. Reprinted figure from \cite{liu_prl_2019}. Copyright CC BY 4.0.} %Reprinted figure with permission from \cite{liu_prl_2019}. Copyright 2019 by the American Physical Society.}
    \label{fig:ar_e2}
\end{figure}

Whilst all evidence suggests a rapid erasure of the $N=34$ shell closure in $Z>20$ isotones, the exotic nature of their $Z<20$ counterparts has made their study difficult. The first $\gamma$-ray spectroscopy of $^{52}$Ar, with the neutron number $\emph{N}$ = 34, was measured using the $^{53}$K$(p,2p)$ one-proton removal reaction at $\sim$210 MeV/u at the RIBF facility. The intensity of the $^{53}$K beam was only 1.0 pps. This measurement was made possible by the high luminosity afforded by the 150-mm long MINOS target, the high photopeak efficiency of the DALI2+ $\gamma$ array, as well as the high primary beam intensities and large acceptance of the SAMURAI spectrometer. With 7-day beam time, 438 counts of $^{52}$Ar were accumulated from $(p,2p)$ reaction. The measured $\gamma$-ray energy spectrum of $^{52}$Ar is presented in Fig. \ref{fig:ar_e2} (a). To quantify the significance level of the observed peaks with relatively low statistics, the likelihood-ratio analysis was performed. The transition at 1656(18) keV was found with a significance level of 5$\sigma$ and the 2295(39) keV $\gamma$ line has a significance of 3$\sigma$. Due to the estimated low $S_{n}$ of $^{52}$Ar and the measured low $S_{n}$ of $^{54}$Ca, two transitions were attributed to direct decay to the ground state. By comparing the measured excitation energies and partial cross sections with theoretical calculations, the 1656(18) state with the highest population is assigned as 2$^{+}_{1}$. As shown in Fig. \ref{fig:ar_e2}(b), the 2$^{+}_{1}$ excitation energy of $^{52}$Ar turns out to be the largest among the Ar isotopes with $\emph{N}$ $>$ 20, even large than $^{46}$Ar with the conventional $\emph{N}$ = 28 shell closure. This result provides the first experimental evidence for the persistence of the $\emph{N}$ = 34 subshell closure beyond $^{54}$Ca. Shell-model calculations with phenomenological and chiral-effective-field-theory interactions both reproduce the measured 2$^{+}_{1}$ systematics of neutron-rich Ar isotopes, and support a $\emph{N}$ = 34 subshell closure in $^{52}$Ar. However, the $\emph{ab initio}$ coupled-cluster calculations reproducing well the neutron-rich Ca data failed to describe the E(2$^{+}_{1}$) in $^{50}$Ar and $^{52}$Ar and called for further development of the calculation.

\subsection{Shell migration towards $\it{N}$ = 40}

%%Possible motivation: N = 40, Island of inversion, New halo....
% FB: I think this bit is rather overlapping with Liliana's contribution... 
The sometimes-presumed $\emph{N}$ = 40 shell closure in the neutron-rich Ca region is associated with the filling of the 1$p$-0$f$ harmonic-oscillator shell, but up to now the closed $fp$-shell configuration was only suggested in $^{68}$Ni. The nuclei $^{66-70}$Fe, $^{64-66}$Cr and $^{62}$Ti have deformed spectra arising from the $fp$-$sdg$ cross shell excitations, which do not reflect an $N=40$ shell closure. 

To obtain more complete experimental data for nuclei around $^{60}$Ca is one of the milestones for the new generation of radioactive-beam facilities such as FRIB~\cite{ostroumov_joi_2020,wei_mpla_2022} and the upgraded RIBF~\cite{ribf_upgrade}. 
Although the spectroscopy of \ts{60}Ca is still not accessible, we can approach \ts{60}Ca following the isotopic chain, performing the first spectroscopy measurements of \ts{55-58}Ca, the results of which can be used to provide well-guided predictions on the structure of \ts{60}Ca.

\begin{figure}[tb]
\centering
\includegraphics[width=\textwidth]{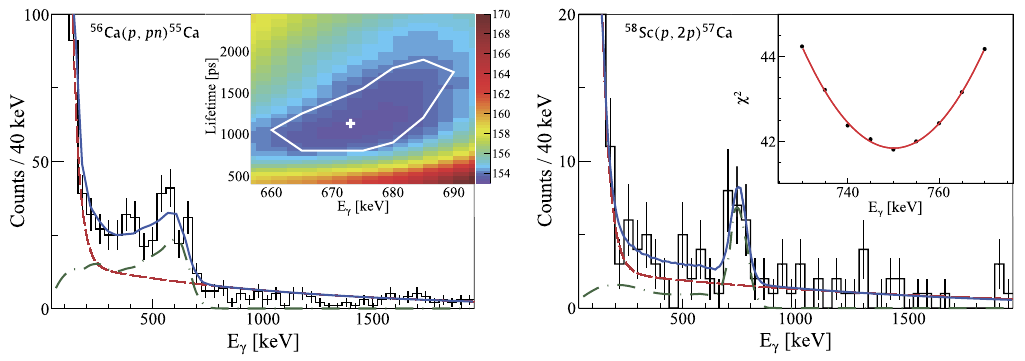}
\caption{\label{fig:Spec5557Ca} Doppler-corrected $\gamma$-ray energy spectra for the $^{56}$Ca$(p,pn)$$^{55}$Ca reaction (left) and $^{58}$Sc$(p,2p)$$^{57}$Ca reaction (right). Reprinted figures from Ref.~\cite{Koiwai2022PLB}. Copyright CC BY 4.0.
}
\end{figure}

%%% 55,57Ca
The excited states of \ts{55,57}Ca were populated by direct proton-induced nucleon removal reactions $^{56}$Ca$(p,pn)$$^{55}$Ca and $^{58}$Sc$(p,2p)$$^{57}$Ca~\cite{Koiwai2022PLB}. The intensities of the $^{56}$Ca and $^{58}$Sc beams were 0.16 and 1.2 pps, respectively. The Doppler-corrected $\gamma$-ray energy spectra are shown in Fig.~\ref{fig:Spec5557Ca}, with one transition observed in each case.
In the \ts{55}Ca spectrum, the broad structure is fitted by a transition at 673(17)\,keV with a lifetime of $\tau = 1130_{-330}^{+520}$\,ps.
This state is interpreted as a neutron $(1p_{1/2})^{-1}(0f_{5/2})^{2}$ configuration and assigned spin parity of $1/2^-$.
The reduced transition probability extracted from the measured lifetime is compatible with a single-particle state.
The population of \ts{55}Ca from the proton removal reaction $^{56}$Sc$(p,2p)$$^{55}$Ca was also studied, but found no bound excited states, which supports the neutron dominance of the 673\,keV state in \ts{55}Ca.
In contrast to \ts{55}Ca, a short-lived state at 751(13)\,keV is observed in the \ts{57}Ca spectrum, suggesting a transition toward many particle-hole dominated configurations beyond $\emph{N}$ = 36.

%%% 56,58Ca
The excited states of \ts{56,58}Ca were both populated following quasi-free one-proton removal reactions from Sc beams. The Sc beam intensities impinging on the MINOS liquid-hydrogen target were 13.6\,particles/s for \ts{57}Sc and 0.3\,particles/s for \ts{59}Sc.
For \ts{56}Ca, a clear $\gamma$-ray transition was observed at 1456(12)\,keV and tentatively assigned as the decay from the $2^+_1$ state to the ground state.
The cross section populating to the $2^+_1$ state was measured to be 0.43(4)\,mb.
For \ts{58}Ca, due to the limited statistics, the structure likely associated with the decay of the $2^+_1$ state observed at 1115(34)\,keV could only be assigned a significance of 2.8$\sigma$, but with a measured cross section of 0.47(19)\,mb compatible with the one from \ts{56}Ca and inline with prediction.

\begin{figure}[tb]
\centering
\includegraphics[scale=0.9]{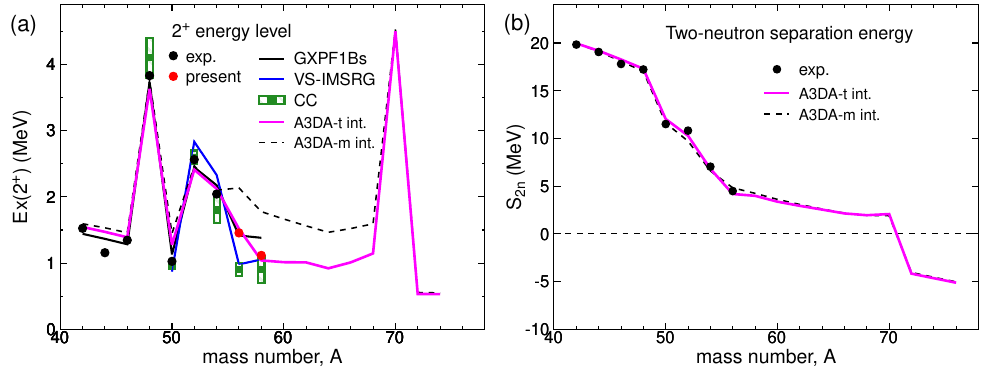}
\caption{\label{fig:CaE2S2n} Comparison of theoretically calculated $E(2^+_1)$ and $S_{2n}$ with experimental data.
(a) $E(2^+_1)$ systematics in even-even Ca isotopes
confronted with contrasting theoretical approaches: The shell model using the GXPF1Bs, A3DA-m and A3DA-t 
Hamiltonian, the VS-IMSRG method, and CC calculations.
(b) $S_{2n}$ systematics in even-even Ca isotopes.
Adapted figures from Ref.~\cite{Chen_2023_PLB}. Copyright CC BY 4.0.}
\end{figure}

The $E(2^+_1)$ systematics in even-even Ca isotopes are shown in Fig.~\ref{fig:CaE2S2n}(a), and compared to conventional shell-model calculations with the GXPF1Bs Hamiltonian in the model space of the full $pf$ shell, and two $\emph{ab initio}$ approaches VS-IMSRG and CC, both employing the chiral interaction 1.8/2.0\,(EM).
The new experimental data exhibit a clear decrease of $E(2^+_1)$ from $\emph{N}$ = 36 to 38, which could not be reproduced by any of the calculations. If the $\emph{N}$ = 40 shell gap is preserved in the Ca isotopes, the $0f_{5/2}$ orbital should be isolated from the higher orbitals. This implies that the $\emph{N}$ = 36 and $\emph{N}$ = 38 fillings correspond to a two-particle system and a two-hole system of the $0f_{5/2}$ orbital, respectively. The $E(2^+_1)$ values are expected to be similar between $\emph{N}$ = 36 and $\emph{N}$ = 38 as a consequence of particle-hole symmetry, regardless of the single-particle energies employed. Therefore, the observed $E(2^+_1)$ decrease suggests a non-isolated $0f_{5/2}$ orbital since as its occupancy increases so does its mixing with higher-lying orbitals, causing the reduction of energy.

Another shell-model calculation was performed using A3DA-m interaction in a model space comprising the full $pf$ shell, the $0g_{9/2}$, and the $1d_{5/2}$ orbitals. As shown in Fig.~\ref{fig:CaE2S2n}(a), this calculation describes well the $E(2^+_1)$ values up to $\emph{N}$ = 34, but deviates for $\emph{N}$ = 36 and 38. However, the energy lowering from $\emph{N}$ = 36 to 38 is well reproduced.
The A3DA-m interaction was revised by varying only the two-body-matrix-elements (TBME) to better reproduce the experimental $E(2^+_1)$ values. The revised interaction, named A3DA-t, predicts a non-doubly magic \ts{60}Ca and the dripline of Ca isotopes being at \ts{70}Ca or even beyond (Fig.~\ref{fig:CaE2S2n}(b)).

\section{Perspectives}
Direct reactions with thick LH$_2$ targets at the RIBF have provided a wealth of experimental feedback on the structures of extremely exotic $Z\approx20$ nuclei over the last 10 years. However, this leads to new and crucial questions that require further experimental and theoretical developments. To further verify, for instance, whether the extended $\emph{p}$-wave neutron orbital observed in the $^{52}$Ca$(p,pn)$$^{51}$Ca reaction \cite{Enciu2022} is responsible for the anomalous charge and matter radius evolution in neutron-rich Ca isotopes beyond $\emph{N}$ = 28, interaction and charge-changing cross section measurements of $^{52-56}$Ca would be highly valuable. Such experiments are feasible with current RIBF intensities, and would test the expectation of an abrupt change of radius at $\emph{N}$ = 34 following the complete filling of the p$_{3/2}$ and p$_{1/2}$ orbitals. From a theoretical standpoint, the evolution of the charge radii in Ca and K isotopes remains a challenge to ab initio and effective theories \cite{Koszorus2021}.

Within this special issue, a dedicated article \cite{tanaka-ptep} on the future devices is presented, here, discussions will focus primarily on the physics cases these devices allow access to.

%STRASSE: intruder bands of 54Ca;
The location of first-excited $0^+$ states is of great interest for magic nuclei, as they provide feedback on valence orbitals' correlations with higher-lying shells. However, if the $0^+_2$ state is the first-excited state, its decay often cannot be measured through in-beam $\gamma$-ray spectroscopy. Alternatively, the population of such states can be accessed using the missing-mass technique. For $(p, 2p)$ and $(p, 3p)$ quasi-free scattering reactions with the LH$_2$ targets, the newly constructed STRASSE and CATANA array~\cite{liu_epja_2023} at the RIBF will provide an optimized setup for missing-mass spectroscopy, with an accepted experiment aiming to search for the $0^+_2$ state in \ts{54}Ca~\cite{liu_RIBF_proposal}. This measurement would elucidate the shell structure built atop the $N=34$ shell closure of \ts{54}Ca. 

%Whilst the $\gamma$-ray spectroscopy utilizing the DALI2+ array features prominently and suits well the studies reported in this article, high-resolution $\gamma$-ray spectroscopy elucidate features inaccessible to moderate energy resolution studies. Of particular note is the ability to measure the lifetimes of excited states down to a few picoseconds. HiCARI~\cite{wimmer_rapr_2021}, a recent initiative to introduce high-resolution in-beam $\gamma$-ray spectroscopy at the RIBF, has been applied to measure the lifetime of the $5/2^-$ state in \ts{53}Ca~\cite{chen_2024_53Ca}. 

Whilst the $\gamma$-ray spectroscopy utilizing the DALI2+ array suits well the studies reported in this article, high-resolution $\gamma$-ray spectroscopy elucidates features of exotic nuclei inaccessible to moderate energy-resolution measurements. Of particular note is the ability to measure the lifetimes of excited states down to a few picoseconds. HiCARI~\cite{wimmer_rapr_2021}, a recent initiative to introduce high-resolution in-beam $\gamma$-ray spectroscopy at the RIBF, has been applied to measure the lifetime of the $5/2^-$ state in \ts{53}Ca~\cite{chen_2024_53Ca}. Future, dedicated germanium-based tracking arrays will significantly enhance the capability to measure lifetimes, thereby extracting reduced transition probabilities. This will complement the measured energy levels in constraining theoretical models.

%Future, dedicated germanium-based tracking arrays will greatly expand the capability to measure lifetimes to extract reduced transition probabilities to complement the measured energy levels in constraining theoretical models. 
%In the article dedicated to the detector development of this topical issue are detailed examples of such dedicated arrays, and the plans to implement one at the RIBF. 

For the $\gamma$-ray spectroscopy of the most exotic nuclei, germanium-based detector array is of limited efficacy owing to their moderate efficiency and peak-to-total ratios, especially for high-energy transitions. As such, detectors based on a material with large stopping power, that is, a high-effective $Z$, are required. To this end, the HYPATIA array is being developed~\cite{hypatia}. Its substantial peak-to-total ratio will enable the spectroscopy of very weakly-produced isotopes. An example measurement will be the spectroscopy of \ts{60}Ca, which would provide definitive evidence of its magicity. Moreover, such an array would be well-suited to determine with high confidence the transitions in \ts{58}Ca, as well as the spectroscopy of nuclei southeast of \ts{54}Ca. 

%Readers are again directed to the detector development article of this special issue for detailed discussion. 

\section{Concluding remarks}
A substantial body of experimental evidence supports the emergence of $\emph{N}$ = 32 and 34 as non-canonical magic numbers. In this work, we provide a detailed overview of the key measurements conducted at the RIBF with the LH$_2$ target, pertaining to the study of shell migration at $\emph{N}$ = 32, 34 and towards $\emph{N}$ = 40 in Ar, K and Ca isotopes.  
%A detailed overview was given of key results from the work conducted over the past decade using LH$_2$ targets at the RIBF pertaining to the $\emph{N}$ = 32, 34 shell closures and beyond in Ar, K, Ca, and Sc isotopes. 
The results prove the robust double magicity of $^{52,54}$Ca, support the persistence of the $\emph{N}$ = 34 subshell closure below $Z = 20$, and provide an indication for the disappearance of $\emph{N}$ = 40 subshell closure in $^{60}$Ca. With the state-of-the-art theoretical calculations, these exotic phenomenon was found to be linked with the tensor forces and 3N forces. The experimental data in neutron-rich Ca region offer stringent tests for calculations, especially the \emph{ab initio} models. Looking to the future, the development of complementary detection devices, as well as order-of-magnitude beam intensity improvements at the RIBF, will allow the observation of intruder bands that cannot decay via $\gamma$-ray emission, provide access to lifetimes of excited states down to the picosecond region, and definitive spectroscopy up to \ts{60}Ca. These advancements will provide a complete picture of the shell evolution in this key region of the nuclear chart and provide confidence in calculations of nuclei that will remain beyond experimental access for the foreseeable future.

\section*{Acknowledgment}
We thank M. Enciu, J. Holt, Y. Utsuno, Y.L. Sun for the discussions and their contributions to the workshop held in York. H.N.L. is supported by the National Key R\&D Program of China (Grant No. 2024YFE0102800, No. 2023YFA1606403), the National Natural Science Foundation of China (Grant No. 12375111), and the Fundamental Research Funds for the Central Universities of China. S.C. acknowledges support from the UK STFC and the Royal Society. 

\raggedbottom

\bibliographystyle{ptephy}
\bibliography{bibliography}
%\begin{thebibliography}{99}
%\bibitem{Meyer} M. Mayer and J. H. D. Jensen, {\it Elementary Theory of Nuclear Shell Structure} (Wiley, New York, 1955).
%\bibitem{Otsuka2020} T. Otsuka, A. Gade, O. Sorlin, T. Suzuki, and Y. Utsuno, Rev. Mod. Phys. {\bf 92}, 015002 (2020).
%\bibitem{Nowacki2021} F. Nowacki, A. Poves, and A. Obertelli, Prog. Part. Nucl. Phys. {\bf 120}, 103866 (2021).
%\bibitem{Bastin2007} B. Bastin, S. Grevy, D. Sohler, O. Sorlin, Z. Dombradi, N.L. Achouri {\it et al.}, Phys. Rev. Lett. {\bf 99}, 022503 (2007).
%\bibitem{Navin2000} A. Navin, D. W. Anthony, T. Aumann, T. Baumann, D. Bazin, Y. Blumenfeld {\it et al.}, Phys. Rev. Lett. {\bf 85}, 266 (2000).
%\bibitem{Thibault1975} C. Thibault, R. Klapisch, C. Rigaud, A. M. Poskanzer, R. Prieels, L. Lessard, and W. Reisdorf, Phys. Rev. C {\bf 12}, 644 (1975).
%\bibitem{Guillemaud-Mueller1984} D. Guillemaud-Mueller, C. Detraz, M. Langevin, and F. Naulin {\it et al.}, Nucl. Phys. A {\bf 426}, 37 (1984).
%\bibitem{Motobayashi1995} T. Motobayashi, Y. Ikedaa, Y. Ando, K. Ieki, M. Inoue, N. Iwasa {\it et al.}, Phys. Lett. B, {\bf 346}, 9 (1995).
%\bibitem{Steppenbeck2015} D. Steppenbeck, S. Takeuchi, N. Aoi, P. Doornenbal, M. Matsushita, H. Wang {\it et al.}, Phys. Rev. Lett. {\bf 114}, 252501 (2015).
%\bibitem{Huck1985} A. Huck, G. Klotz, A. Knipper, C. Miehe, C. Richard-Serre, G. Walter, A. Poves, H. L. Ravn, G. Marguier {\it et al.}, Phys. Rev. C {\bf 31}, 2226 (1985).

%\end{thebibliography}

\end{document}